\documentclass[aps,prl,twocolumn,showpacs,groupedaddress]{revtex4}  
\usepackage{graphicx}  
\usepackage{dcolumn}   
\usepackage{bm}        
\usepackage{amssymb}   
\usepackage{xspace}

\def\met{\mbox{${\hbox{$E$\kern-0.6em\lower-.1ex\hbox{/}}}_T$}}
\newcommand{\DO}{D\O\xspace}

\begin{document}

\title{First study of the radiation-amplitude zero in $W\gamma$
  production and limits on anomalous $WW\gamma$ couplings at
  $\sqrt{s}=1.96$~TeV}

%
\author{V.M.~Abazov$^{36}$}
\author{B.~Abbott$^{75}$}
\author{M.~Abolins$^{65}$}
\author{B.S.~Acharya$^{29}$}
\author{M.~Adams$^{51}$}
\author{T.~Adams$^{49}$}
\author{E.~Aguilo$^{6}$}
\author{S.H.~Ahn$^{31}$}
\author{M.~Ahsan$^{59}$}
\author{G.D.~Alexeev$^{36}$}
% ...

\affiliation{\vspace{0.1 in}(The D\O\ Collaboration)\vspace{0.1 in}}
% ...
\affiliation{$^{6}$University of Alberta, Edmonton, Alberta, Canada,
                Simon Fraser University, Burnaby, British Columbia, Canada,
                York University, Toronto, Ontario, Canada, and
                McGill University, Montreal, Quebec, Canada}
% ...
\affiliation{$^{29}$Tata Institute of Fundamental Research, Mumbai, India}
% ...
\affiliation{$^{31}$Korea Detector Laboratory, Korea University, Seoul, Korea}
% ...
\affiliation{$^{36}$Joint Institute for Nuclear Research, Dubna, Russia}
% ...
\affiliation{$^{49}$Florida State University, Tallahassee, Florida 32306, USA}
% ...
\affiliation{$^{51}$University of Illinois at Chicago,
                Chicago, Illinois 60607, USA}
% ...
\affiliation{$^{59}$Kansas State University, Manhattan, Kansas 66506, USA}
% ...
\affiliation{$^{65}$Michigan State University,
                East Lansing, Michigan 48824, USA}
% ...
\affiliation{$^{75}$University of Oklahoma, Norman, Oklahoma 73019, USA}
% ...
\date{February 29, 2008}

\begin{abstract}
  We present results from a study of $p \bar{p} \rightarrow W\gamma +
  X$ events utilizing data corresponding to 0.7~fb$^{-1}$ of
  integrated luminosity at $\sqrt{s} = 1.96$~TeV collected by the D0
  detector at the Fermilab Tevatron Collider. We set limits on
  anomalous $WW\gamma$ couplings at the 95\% C.L. The one dimensional
  95\% C.L.\ limits are $0.49 < \kappa_\gamma < 1.51$ and $-0.12 <
  \lambda_\gamma < 0.13$.  We make the first study of the
  charge-signed rapidity difference between the lepton and the photon
  and find it to be indicative of the standard model
  radiation-amplitude zero in the $W\gamma$ system.
\end{abstract}

\pacs{12.15.Ji, 13.40.Em, 13.85.Qk}
\maketitle 

Self-interactions of the electroweak bosons are a consequence of the
$SU(2)_{L} \times U(1)_{Y}$ gauge symmetry of the standard model
(SM). In this Letter, we investigate the $WW\gamma$ vertex by studying
the production of $p \bar{p} \rightarrow W\gamma \rightarrow \ell \nu
\gamma + X$ events where $\ell$ is an electron or a muon. At leading
order (LO), the SM allows $q\bar{q}^\prime \to W\gamma$ production in
which a photon radiates off an incoming quark (initial state
radiation) or is directly produced from the $WW\gamma$ vertex. In the
SM, these two cases involve three amplitudes where each alone violates
unitarity, but together interfere to give a finite cross section. This
interference leads to a radiation-amplitude zero (RAZ) in the angular
distribution of the photon. In this Letter, we set limits on non-SM
$WW\gamma$ couplings and present a first measurement of the
destructive interference indicative of the RAZ in the $W\gamma$
system.

Non-SM $WW\gamma$ couplings will give rise to an increase in the
$W\gamma$ production cross section over the SM prediction,
particularly for energetic photons. CP-conserving couplings may be
parameterized by an effective Lagrangian~\cite{lag, radzero} with two
parameters, $\kappa_\gamma$ and $\lambda_\gamma$, related to the
magnetic dipole and electric quadrupole moments of the $W$ boson. In
the SM, $\kappa_\gamma = 1$ and $\lambda_\gamma = 0$. The effective
Lagrangian with non-SM couplings will violate unitarity at high
energies, and so a form factor with a scale $\Lambda$ is introduced to
modify the coupling parameters with $a_0 \rightarrow a_0 / (1 +
\hat{s}/\Lambda^{2})^{2}$ where $a_0 = \kappa_\gamma, \,
\lambda_\gamma$, and $\sqrt{\hat{s}}$ is the $W\gamma$ invariant
mass. We set $\Lambda$ to 2~TeV~\cite{Lambda}.

A general consequence of gauge theories is that any four-particle tree
amplitude involving one or more massless gauge bosons may be
factorized into a charge dependent part and a spin/polarization
dependent part. The charge dependent part will lead to the amplitude
vanishing at a particular point in phase space. For a $2 \rightarrow
2$ process, as is the case for $W\gamma$, this effect is evident as a
zero in the production amplitude in the angular distribution of the
photon~\cite{radzero}.  The RAZ manifests itself as a dip in the
charge-signed rapidity difference, $Q_\ell\times\Delta y$, between the
photon and the charged lepton from the $W$ boson decay~\cite{CSRD}.
In the massless limit regime, the rapidity difference can be
approximated by the pseudorapidity difference~\cite{measures}, which
can be very precisely measured. The SM predicts that the dip minimum
depends on the quark electric charges and lies at
$Q_\ell\times\Delta\eta \approx -1/3$. In the case of anomalous
couplings the location of the dip minimum does not change, instead the
dip may become more shallow or disappear entirely.

$W\gamma$ production has been studied previously at hadron
colliders~\cite{prevD0}.  The limits set by the most recent previous
\DO\ analysis represented the most stringent constraints on anomalous
$WW\gamma$ couplings obtained by direct observation of $W\gamma$
production. The present analysis uses more than four times as much
data as well as photons in the end-cap calorimeter, and thus has an
increased sensitivity for the study of $Q_\ell\times\Delta\eta$.  The
D0 detector~\cite{d0det} is used in this study to observe $p \bar{p}
\rightarrow \ell \nu \gamma +X \; (\ell = e \text{ or } \mu)$ in
collisions at $\sqrt{s} = 1.96$~TeV at the Fermilab Tevatron
collider. The data samples correspond to integrated luminosities of
$717 \pm 44$~pb$^{-1}$ and $662 \pm 40$~pb$^{-1}$ for the electron and
muon channels, respectively.

Candidate events with the $W$ boson decaying into an electron and a
neutrino are collected with a suite of single electron triggers. The
reconstructed electron is required to be in the central $(|\eta_{\rm
  det}| < 1.1)$ or end-cap $(1.5 < |\eta_{\rm det}| < 2.5)$
calorimeters~\cite{measures}, have transverse energy $E_{T} > 25$~GeV,
be isolated in the calorimeter, have a shower shape consistent with
that of an electromagnetic object, and match a track reconstructed in
the central tracking system. The missing transverse energy, \met, must
exceed 25~GeV. To reduce final state radiation of photons from
leptons, the reconstructed $W$ transverse mass must exceed
50~GeV$/c^{2}$. Furthermore, to suppress background from $Z
\rightarrow ee$ events with an electron misidentified as a photon, the
two-body invariant mass of the electron and photon must be outside the
mass window 87-97~GeV$/c^2$. The optimized window limits are
asymmetric about the $Z$ boson mass because the expected signal will
have more events below the $Z$ boson mass than above it.

Candidate events with the $W$ boson decaying into a muon and a
neutrino are collected with a suite of single muon triggers. The
reconstructed muon is required to be within $|\eta_{\text{det}}| <
1.6$, isolated in the central tracking system and the calorimeter and
be associated with a central track with $p_{T} > 20\text{ GeV}/c$. The
event \met\ must exceed 20~GeV and there must be no additional
isolated tracks with $p_{T} > 15\text{ GeV}/c$ as well as no
additional muons. The muon momentum is measured by the curvature of
the track in the central tracking system.

Photons are identified with the same requirements in both
channels. The photon must have $E_T > 9$~GeV and be in the central
$(|\eta_{\rm det}| < 1.1)$ or end-cap $(1.5 < |\eta_{\rm det}| < 2.5)$
calorimeter. It must be isolated in the calorimeter and tracker, have
a shower shape consistent with that of an electromagnetic object, have
an associated cluster in the preshower detector, and, if in the
central region, project back to a position along the beam axis within
10~cm of the primary vertex. The photon and the lepton must be
separated in $\eta -\phi$ space by $\Delta R = \sqrt{(\Delta \eta)^{2}
  + (\Delta \phi)^{2}} > 0.7$. To further suppress final state
radiation, the three-body transverse mass ($M_{T3}$) of the photon,
lepton, and missing transverse energy must exceed 120~GeV$/c^2$ and
110~GeV$/c^2$ for the electron and muon channels, respectively.

Kinematic and geometric acceptances are determined using Monte Carlo
(MC) events. For the acceptances to be meaningful, they are measured
with respect to reference kinematic requirements of $E^\gamma_T >
9$~GeV, $M_{T3} > 90$~GeV$/c^2$, and $\Delta R > 0.7$ (MC samples were
produced with much looser requirements). A LO simulation~\cite{baur}
of $W\gamma$ production is used, which includes the contributions from
initial and final state radiation as well as the $WW\gamma$ trilinear
vertex. To compensate for the effects of next-to-leading order (NLO)
corrections on the $E^\gamma_T$ spectrum, a NLO MC~\cite{baurNLO} is
used, and an $E^\gamma_T$-dependent $K$-factor is calculated and
applied to the LO spectra. The detector resolutions are applied using
a parameterized simulation.

Electron and muon identification efficiencies are determined with
large $Z \rightarrow e e$ or $Z \rightarrow \mu \mu$ samples from the
data. The photon detection efficiency is determined by the full {\sc
  geant}~\cite{geant} detector simulation and is verified with
$Z\gamma$ data. In these events, the photon is radiated from a final
state lepton and so the three-body mass of the photon and the leptons
should reconstruct the $Z$ boson mass. The reconstruction
efficiency from the {\sc geant} MC is scaled to match the measured
efficiency from the $Z\gamma$ process in data. The acceptance times
efficiency values described here are shown in Table~I.

\begin{table}
  \caption{\label{tab:summary}Summary of event yields. When 
    uncertainties are shown, the first is statistical and the second is systematic.  When only one uncertainty is shown, it is systematic.}
\begin{ruledtabular}
\begin{tabular}{lccc}
                                 & $e\nu\gamma$ channel     & $\mu\nu\gamma$ channel \\ \hline
  Luminosity                     & $720 \pm 44$~pb$^{-1}$    & $660 \pm 40$~pb$^{-1}$ \\
  Acceptance $\times$ efficiency & $0.063 \pm 0.003$        & $0.045 \pm 0.003$ \\ \hline
  $W$ + jet background               & $34 \pm 3.8 \pm 3.1$     & $18 \pm 2.9\pm 1.9$ \\
  $\ell e X$ background          & $17 \pm 2.7 \pm 1.3$     & $2.7\pm 1.3\pm 0.2$ \\
  $W \rightarrow \tau$ background & $1.1 \pm 0.1 \pm 0.1$   & $1.4\pm 0.2\pm 0.1$ \\
  $Z\gamma$ background           & ---                      & $3.8\pm 0.53\pm 0.42$ \\ \hline
  Candidate events                & 180                      & 83\\
  Measured signal & $130 \pm 14 \pm 3.4$   & $57 \pm 8.8 \pm 1.8$ \\
  SM prediction & $120 \pm 12$                          & $77 \pm 9.4$ \\
\end{tabular}
\end{ruledtabular}
\end{table}
  
Backgrounds to $W\gamma$ production include $W$ + jet events where the
jet is misidentified as a photon; ``$\ell e X$'' events with a lepton,
electron, and \met\ where the electron is misidentified as a photon;
$Z\gamma \to \ell\ell\gamma$ events where a lepton is lost; and
$W\gamma \rightarrow \tau\nu\gamma$.  The $W$ + jet background
dominates both channels and is determined from data. The rate at which
a jet is misidentified as a photon is calculated from a large multijet
sample in which the jets under study are required to have a large
fraction of their energy deposited in the electromagnetic layers of
the calorimeter. This rate is calculated as a function of $E_T$ and
$\eta_{\rm det}$. The rate is then applied to a normalization sample
of $W$ + jet events where the jets satisfy the same criteria as in the
multijet sample. To determine the $\ell e X$ background, the track
isolation requirement is removed from the photon and a matched track
is required. The measured tracking efficiencies are then used to
estimate this background contribution. The $Z\gamma$ and $W\gamma
\rightarrow \tau \nu \gamma \rightarrow e (\mu) \nu \nu \gamma$
backgrounds are estimated from MC. The $Q_\ell\times\Delta\eta$
distribution of the total background lacks any statistically
significant structure.  A summary of the background estimates and the
observed $W\gamma$ candidate events are shown in Table I.

Since the observed event yields are consistent with the SM
predictions, limits on anomalous $WW\gamma$ trilinear couplings are
determined using the combined $E^\gamma_T$ spectrum from both channels
(Fig.~1).  Limits are set by generating $E^\gamma_T$ spectra for
different values of the coupling parameters $\kappa_\gamma$ and
$\lambda_\gamma$, and then calculating the likelihood they represent the
data. The 95\%~C.L.\ limit contour is found numerically by integrating
the likelihood surface and finding the minimum contour that represents
95\% of the volume. One-dimensional 95\% C.L.\ limits are calculated by
setting one coupling parameter to the SM value and allowing the other
to vary. These limits, shown in Fig.~2, are $0.49 < \kappa_\gamma <
1.51$ and $-0.12 < \lambda_\gamma < 0.13$.

\begin{figure}
\includegraphics[width=\linewidth]{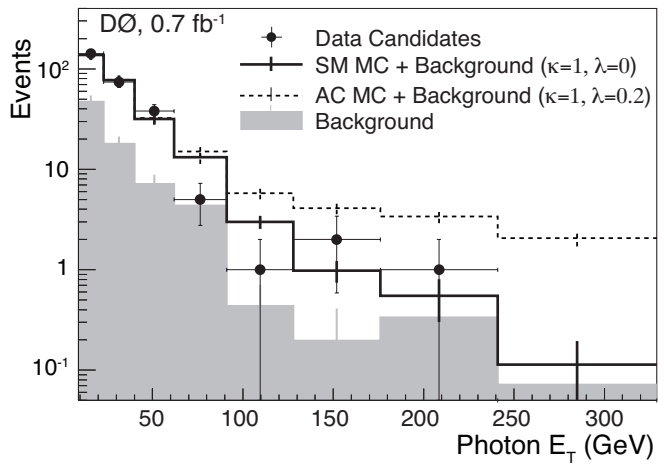}
\caption{\label{fig:phoEt} The photon transverse energy spectra for
  the SM (solid line), an anomalous coupling (AC) point (dashed line),
  combined electron and muon channel data candidates (black points),
  and the background estimate (shaded histogram). Uncertainties are
  shown as error bars on the points, lines, and histograms. The last
  bin includes overflows. }
\end{figure}

\begin{figure}
\includegraphics[width=\linewidth]{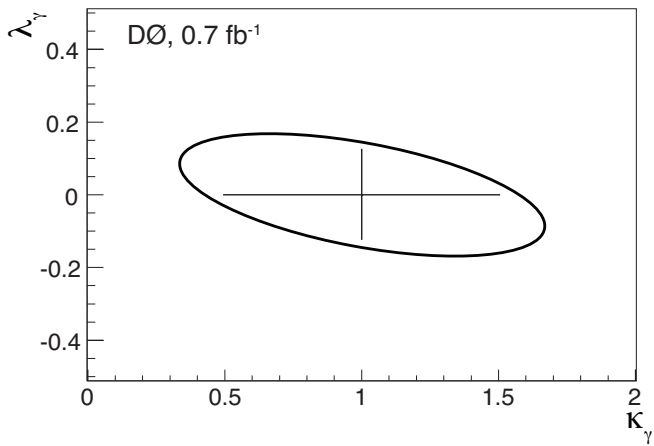}
\caption{\label{fig:limit}The ellipse is the 95\% C.L.\ limit contour in
  $\kappa_\gamma -\lambda_\gamma$ space. One-dimensional 95\% C.L.\ limits are shown as
the horizontal and vertical bars.}
\end{figure}

The background-subtracted $Q_\ell\times\Delta\eta$ distribution for
the combined electron and muon channels is shown in Fig.~3. To perform
a statistical test for the presence of a dip, the distribution is
divided into two bins whose edges are determined by the
$Q_\ell\times\Delta\eta$ distribution generated in SM Monte Carlo. The
bins are chosen to be adjacent and of equal width such that one
samples the majority of events in the dip and the other samples the
smaller of the local maxima (see the inset in Fig.~4). We define a
test statistic $R$ to be the ratio of the integral number of events in
the dip bin to the integral number of events in the maximum bin.  This
ratio will be at least one if there is no dip (unimodal distribution),
and less than one if there is a dip. For the combined
background-subtracted data $Q_\ell\times\Delta\eta$, this ratio test
gives a value of 0.64.

\begin{figure}
\includegraphics[width=\linewidth]{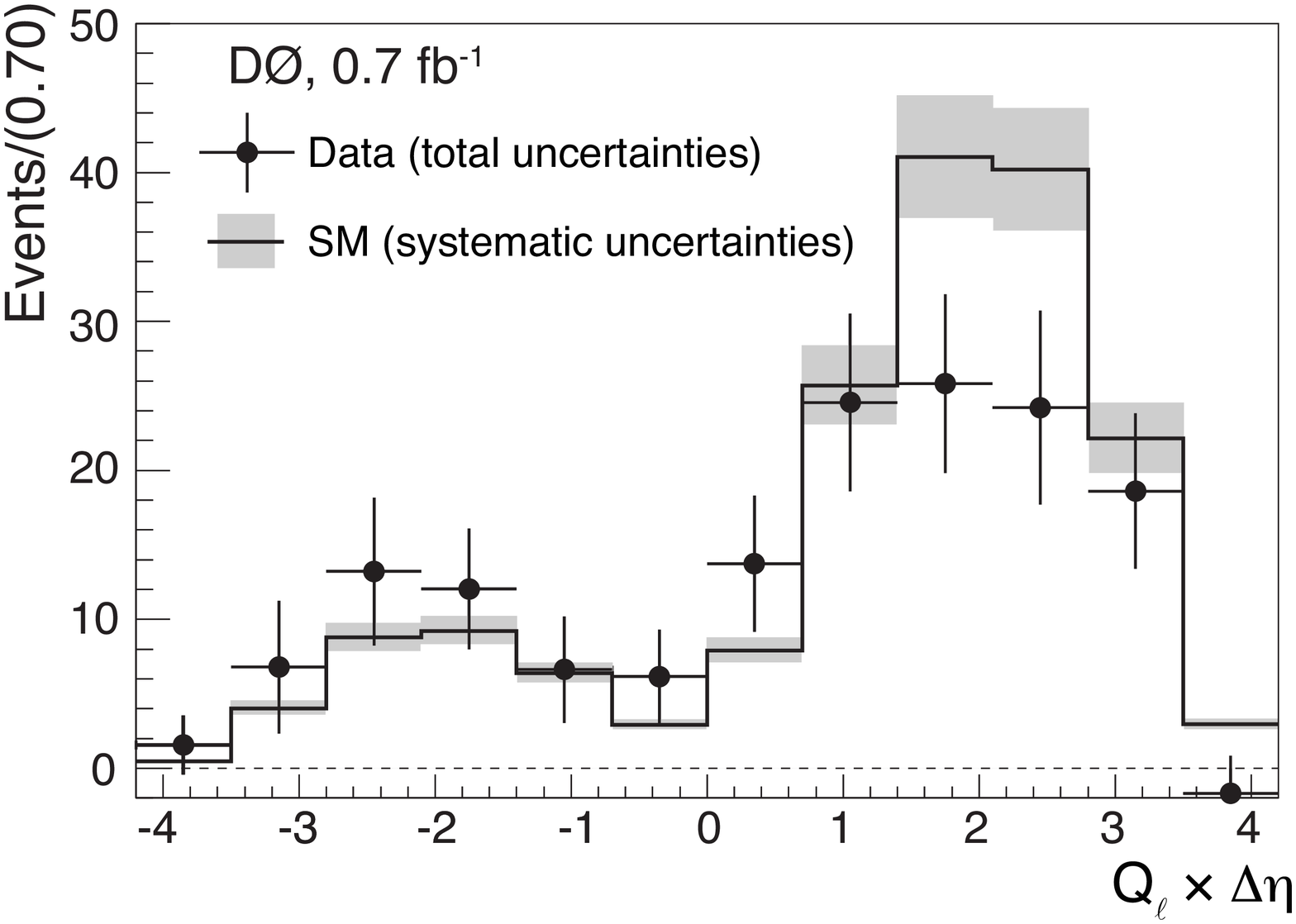}
\caption{\label{fig:csrd} The background-subtracted charge-signed
  rapidity difference for the combined electron and muon channels. The
  black points and error bars represent background-subtracted data
  with its associated uncertainties (statistical and from the
  subtraction procedure), and the shaded areas are the systematic
  uncertainties on the SM prediction (including on efficiencies and
  acceptances). The solid line is the distribution from the SM. A
  $\chi^2$ test comparing the data and SM using the full covariance
  matrix yields 17 for 12 degrees of freedom, indicating good
  agreement.}
\end{figure}

We first compare this observed $R$ value from the data to an ensemble
of $10^4$ MC SM pseudo-experiments where all statistical and
systematic fluctuations are included. For the SM, 28\% of the
experiments have a ratio of 0.64 or greater. In order to evaluate the
significance of the observed data $R$ value, we select an anomalous
coupling value which provides a $Q_\ell\times\Delta\eta$ distribution
that minimally exhibits no dip --- the minimal unimodal hypothesis
(MUH).  Minimal specifically means a class of distributions on the
boundary of bimodal and unimodal distributions. The distribution
chosen here corresponds to $\kappa_{\gamma}=0$, $\lambda_{\gamma} =
-1$ (zero magnetic dipole moment of the $W$ boson). Anomalous
couplings increase the event yield as well, but since we are only
concerned with the distribution shape, we normalize this distribution
to the number of events predicted by the SM. For this MUH case, only
45 experiments out of $10^4$ have an $R$ value of 0.64 or smaller due
to a random fluctuation. These distributions are shown in Fig.~4. If
transformed into a Gaussian significance, this probability corresponds
to $2.6\sigma$. This result is the first study of the
$Q_\ell\times\Delta\eta$ distribution and is indicative of the RAZ in
$W\gamma$ production.

\begin{figure}
\includegraphics[width=\linewidth]{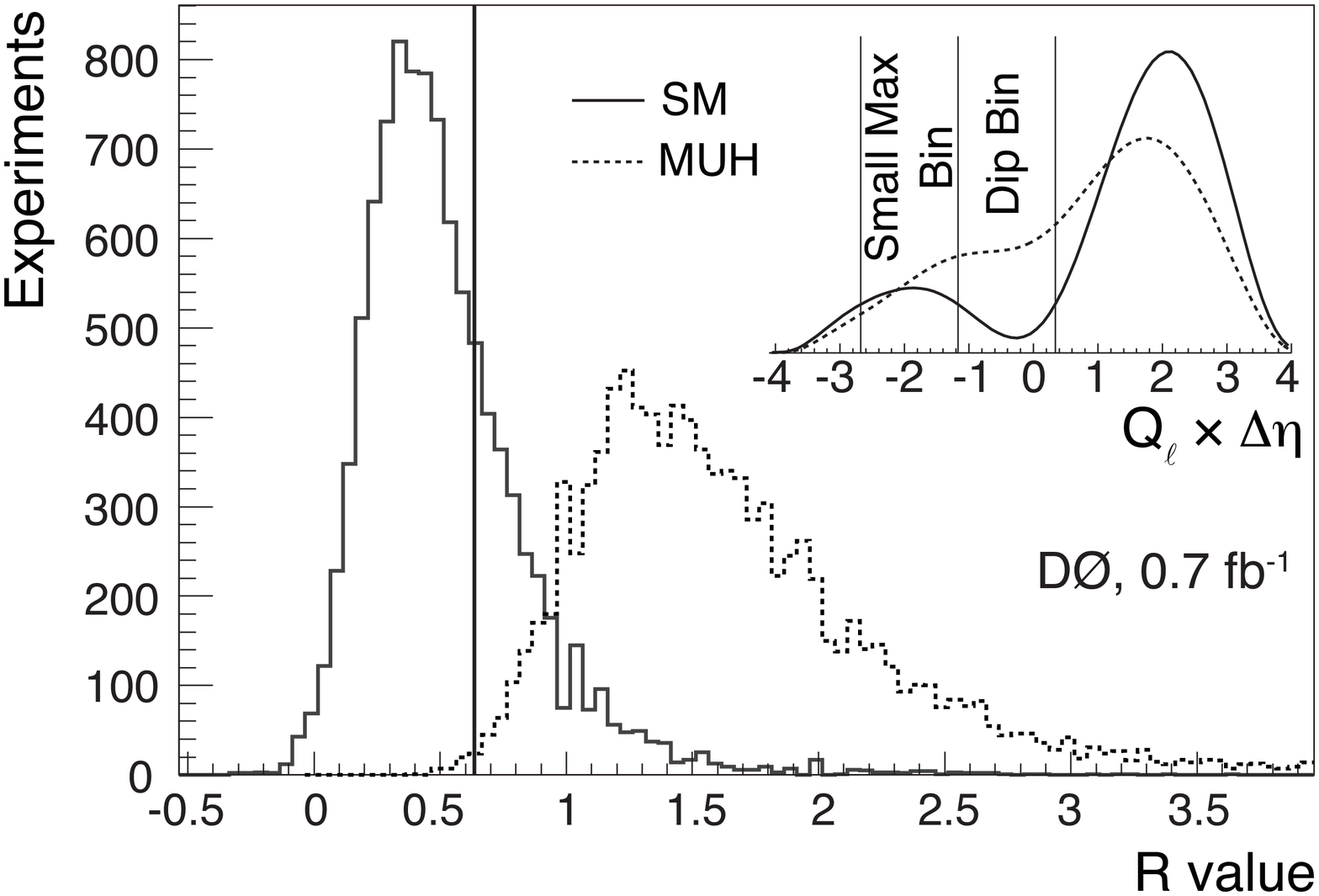}
\caption{\label{fig:rTest} Distributions of the $R$-test statistic for
  the SM ensembles (solid line) and the MUH ensembles (dashed
  line). The vertical line indicates the measured value from the
  data. The inset plot indicates the positions of the two bins used
  for the $R$-test as determined by a fit to the SM
  $Q_\ell\times\Delta\eta$ distribution (solid line). For comparison,
  a fit to the MUH $Q_\ell\times\Delta\eta$ distribution is shown as
  the dashed line.}
\end{figure}

In summary, we have studied $W\gamma$ production and set 95\% C.L.\
limits on anomalous trilinear gauge couplings at $0.49 < \kappa_\gamma
< 1.51$ and $-0.12 < \lambda_\gamma < 0.13$. These limits are the most
stringent set at a hadron collider for this final state. We also
performed the first study of the radiation-amplitude zero in the
charge-signed rapidity difference between the lepton and the
photon. The probability that this measurement would arise from a
minimal unimodal hypothesis is smaller than $(4.5 \pm 0.7) \times
10^{-3}$ and is indicative of the radiation-amplitude zero in
$W\gamma$ production.

We thank Prof.\ David W.\ Scott, Department of Statistics, Rice
University. 
We thank the staffs at Fermilab and collaborating institutions, 
and acknowledge support from the 
DOE and NSF (USA);
CEA and CNRS/IN2P3 (France);
FASI, Rosatom and RFBR (Russia);
CNPq, FAPERJ, FAPESP and FUNDUNESP (Brazil);
DAE and DST (India);
Colciencias (Colombia);
CONACyT (Mexico);
KRF and KOSEF (Korea);
CONICET and UBACyT (Argentina);
FOM (The Netherlands);
STFC (United Kingdom);
MSMT and GACR (Czech Republic);
CRC Program, CFI, NSERC and WestGrid Project (Canada);
BMBF and DFG (Germany);
SFI (Ireland);
The Swedish Research Council (Sweden);
CAS and CNSF (China);
and the
Alexander von Humboldt Foundation.
%


\begin{thebibliography}{99}
\bibitem{measures} D0 uses a cylindrical coordinate system with the
    $z$ axis running along the beam axis. Angles $\theta$ and $\phi$
    are the polar and azimuthal angles, respectively. Pseudorapidity
    is defined as $\eta = -\ln \left[ \tan(\theta/2) \right]$ where
    $\theta$ is measured with respect to the primary vertex. In the
    massless limit, $\eta$ is equivalent to the rapidity $y = (1/2)
    \ln \left[ (E+p_z)/(E-p_z) \right]$. $\eta_{\rm det}$ is the
    pseudorapidity measured with respect to the center of the
    detector.
\end{thebibliography}
\end{document}